\documentclass{PoS}
\usepackage{amssymb,amsmath}
\usepackage{upgreek}
\usepackage[permil]{overpic}
\usepackage{pict2e}
\usepackage{booktabs}
\usepackage{caption}
\usepackage{subcaption}

\newcommand{\pT}          {$p_\text{T}$}
\newcommand{\NIEL}         {1~MeV~$\text{n}_{\text{eq}}\text{/cm}^2$}

\title{Upgrade of the Inner Tracking System of ALICE}

\ShortTitle{Upgrade of the Inner Tracking System of ALICE}

\author{\speaker{Monika Kofarago}%
        \thanks{on behalf of the ALICE collaboration.}\\
        European Organization for Nuclear Research (CERN), Switzerland and\\
        Utrecht University, Netherlands\\
        E-mail: \email{monika.kofarago@cern.ch}}

\abstract{The upgrade of the Inner Tracking System (ITS) of ALICE is planned for the second long shutdown of the LHC in 2019--2020. The ALICE physics program after the shutdown requires the ITS to have improved tracking capabilities and improved impact parameter resolution at very low transverse momentum, as well as a substantial increase in the readout rate. To fulfill these requirements the current ITS will be replaced by seven layers of Monolithic Active Pixel Sensors. The new detector will be moved as close as 23 mm to the interaction point and will have a significantly reduced material budget. Several prototypes of the sensor have been developed to test different aspects of the sensor design including prototypes with analog and digital readout, as well as small and final-size sensors. These prototypes have been thoroughly characterized both in laboratory tests and at test beam facilities including studies on the radiation hardness of the sensors. This contribution gives an overview of the current status of the research and development with a focus on the pixel sensors and the characterization of the latest prototypes. }

\FullConference{24th International Workshop on Vertex Detectors\\
		 1-5 June 2015\\
		 Santa Fe, New Mexico, USA}

\begin{document}

%===================================%
\section{Upgrade of ALICE}
  The ALICE experiment was designed to study the properties of the quark-gluon plasma created in heavy-ion collisions at the LHC. Since the start of the LHC, ALICE has confirmed the existence of the quark-gluon plasma and extended the precision and kinematic reach of many significant probes measured previously at RHIC and at the SPS. After the second long shut down of the LHC in 2019--2020, ALICE will focus on the measurement of heavy-flavor hadrons, quarkonia and low-mass dileptons at low transverse momenta \cite{ref:ALICE_LOI}. This requires high-precision measurements at low \pT{} resulting in the need of recording large samples of minimum-bias events. In order to achieve this, ALICE is planning to read out all Pb-Pb collisions delivered by the LHC which corresponds to a maximum collision rate of $50$ kHz. The upgraded detector will record an integrated luminosity of $10\text{ nb}^{-1}$ in Pb-Pb collisions which means a factor $100$ compared to the data that will be recorded by 2019.

  To cope with the physics requirements of ALICE several upgrades will take place during the second long shut down of the LHC. Two new detectors will be installed: a high resolution, low material Inner Tracking System (ITS) and a Muon Forward Telescope (MFT). They will be accompanied by a new beampipe with a smaller radius which will allow the first layer of the ITS to be moved closer to the interaction point resulting in better impact-parameter determination. The readout chambers of the Time Projection Chamber (TPC) will be replaced by Gas Electron Multiplier (GEM) detectors and its readout electronics will be upgraded together with the readout electronics of the Transition Radiation Detector (TRD), the Time Of Flight (TOF) detector and the Muon Spectrometer. The forward trigger detectors, the online system and the offline reconstruction and analysis framework will also be upgraded \cite{ref:ALICE_LOI}.

%===================================%
\section{Upgrade of the Inner Tracking System}
\begin{figure}[!htbp]
  \begin{center}
    \begin{overpic}[width=0.55\textwidth]{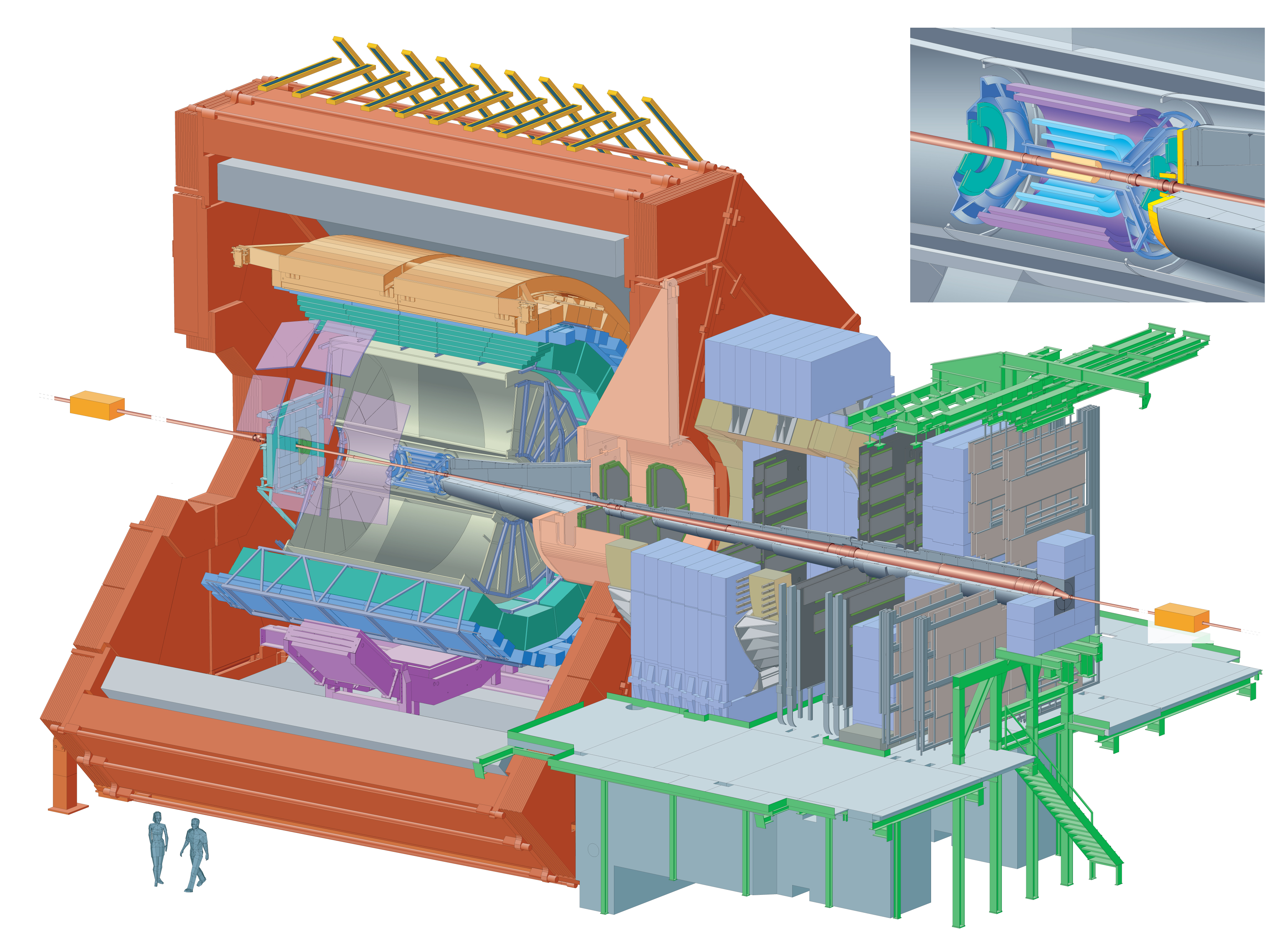}
      \put(560,670){\Large ITS}
      \put(1015,650){\large Strip}
      \put(1010,562){\large Drift}
      \put(960,450){\large Pixel}
      \linethickness{1.1pt}
      \put(600,660){\vector(-1,-1.05){270}}
      \put(610,660){\vector(1,-0.3){130}}
      \put(1000,660){\vector(-1,-0.1){140}}
      \put(1000,575){\vector(-1,0.3){140}}
      \put(970,490){\vector(-1,0.85){130}}
    \end{overpic}
    \caption{Current setup of the ALICE detector with the ITS highlighted.}
    \label{fig:alice}
  \end{center}
\end{figure}
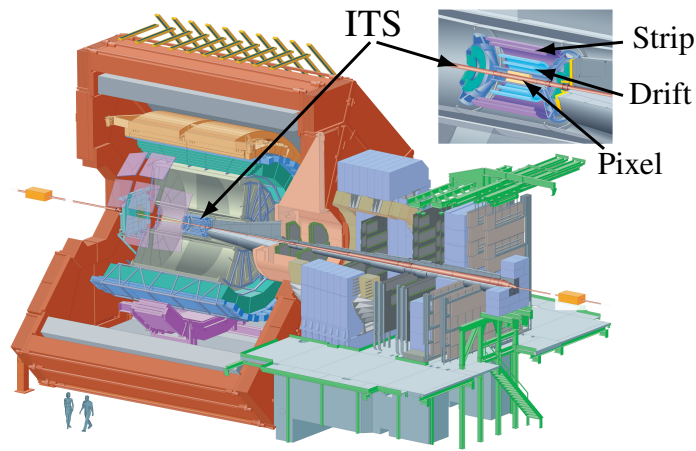

  The Inner Tracking System is situated at the center of ALICE closest to the interaction point (Fig. \ref{fig:alice}). The data taking by the ITS before the first long shut down of the LHC has been successful and data taking has been already resumed since the restart of the accelerator \cite{ref:currentITSTalk}. 

  However, the current detector is not suited for the physics requirements of ALICE after the second long shut down of the LHC, and an upgrade is necessary to improve the physics performance of ALICE in several ways. The impact parameter resolution at \pT{} $=500$ MeV/c will be improved by a factor $5$ in the direction of the beam and a factor $3$ in the transverse direction which will be achieved by different steps. Firstly, the smaller beampipe will allow the first layer of the ITS to be moved as close as $23$ mm to the interaction point instead of the current $39$ mm. Secondly, the material budget of the detector will be reduce from around $1.14\%\text{ X}_0$ to around $0.3\%\text{ X}_0$ for the three innermost layers and finally the pixel size will be reduced to about $30 ~\upmu\text{m} \times 30 ~\upmu\text{m}$ from the current $50~\upmu\text{m} \times 425 ~\upmu\text{m}$. The upgraded detector will also have an improved tracking efficiency and \pT{} resolution at low \pT. To achieve this, the current six layers of hybrid pixel, strip and drift detectors will be replaced by seven layers of monolithic pixel detectors.

  Right now ALICE is limited by the current ITS in the readout speed to $1$ kHz. After the upgrade, ALICE is planning to read out all Pb-Pb collisions at the maximum rate of the LHC: thanks to the design of the new detector a readout rate of more than $10^5$ Hz in Pb-Pb events and several $10^5$ Hz in pp events will be possible. The requirement to remove and reinsert the detector during the yearly shut down of the LHC places further strong constraints on the design of the mechanical support of the detector.

\begin{figure}[!htbp]
  \begin{center}
    \begin{overpic}[width=0.55\textwidth]{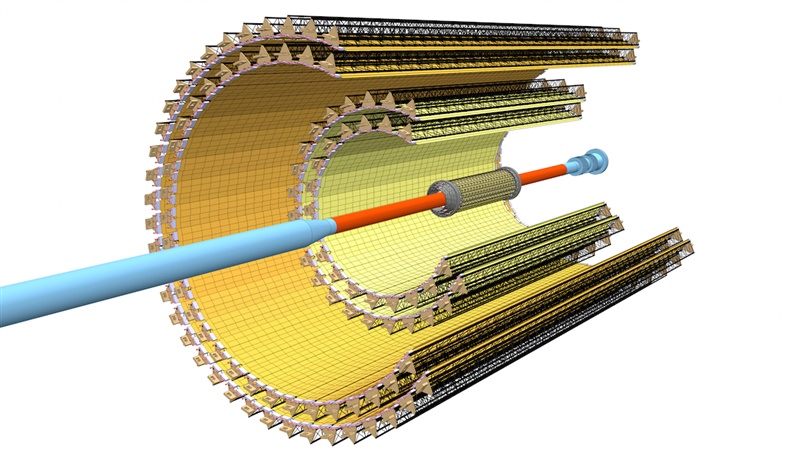}
      \put(   0,100){\large{Beam pipe}}
      \put(   -170,340){\large{Outer Barrel}}
      \put( 750,400){\large{Inner Barrel}}
      \linethickness{1.5pt}
      \put( 747,398){\vector(-2,-1){95}}
      \put( 90,340){\vector(2,-1){110}}
      \put( 90,350){\vector(2,-0.2){290}}
    \end{overpic}
    \caption{Layout of the layers of the upgraded ITS. Figure is taken from \cite{ref:TDR}.}
    \label{fig:ITS_layout}
  \end{center}
\end{figure}
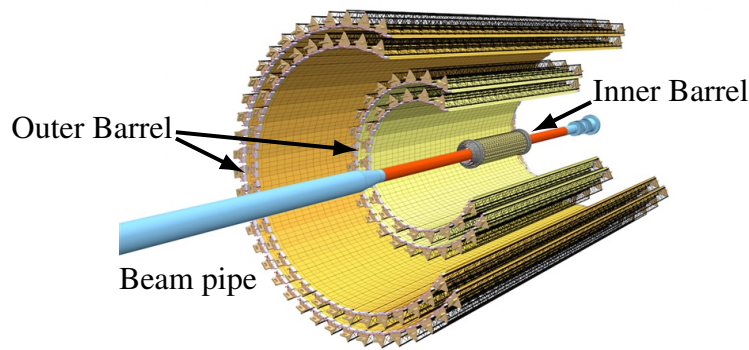

\begin{figure}[!t]
  \begin{subfigure}[b]{0.49\textwidth}
    \begin{overpic}[width=0.85\textwidth]{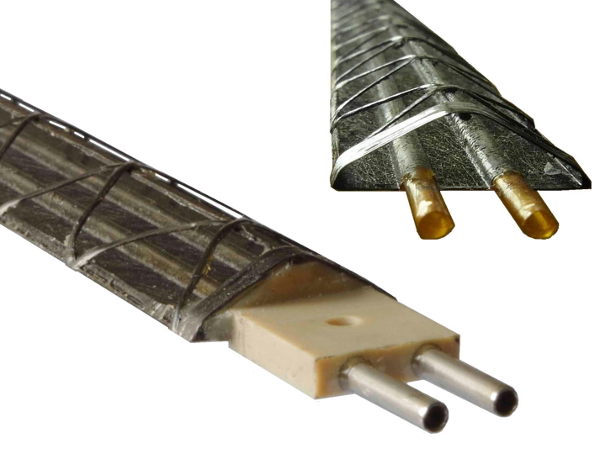}
    \end{overpic}
  \caption{Prototype of the Inner Barrel stave.}
  \end{subfigure}
  \begin{subfigure}[b]{0.49\textwidth}
    \begin{overpic}[width=0.8\textwidth]{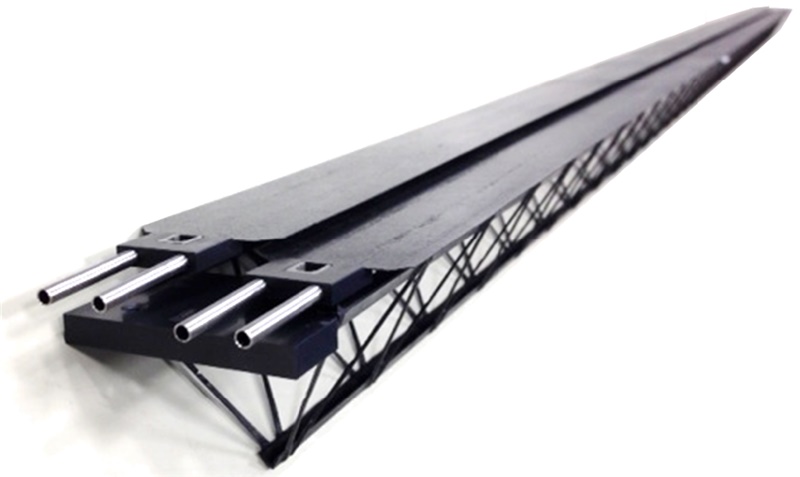}
    \end{overpic}
  \caption{Prototype of the Outer Barrel stave.}
  \end{subfigure}
  \caption{Stave prototypes.}
  \label{fig:staves}
\end{figure}

  The seven layers of the new ITS cover an area of around $10\text{ m}^2$ and will consists of around $1.25 \times 10^{10}$~\mbox{pixels}. The layers will be grouped according to the following: the three innermost layers make up the Inner Barrel and the four outermost layers the Outer Barrel. The radial arrangement of the layers can be seen in Fig. \ref{fig:ITS_layout}. All the layers will be segmented into elements called staves which extend the full length of the detector in the direction of the beam. A stave is the smallest operable unit of the detector and has a slightly different design in the case of the Inner and Outer Barrel. The first prototypes of both the Inner and Outer Barrel staves can be seen in Fig. \ref{fig:staves}.

  Since the distance of the layers from the interaction point vary from $23$ mm to $400$ mm, the working environment and the requirements are very different on the innermost and the outermost layers. The requirements for the different layers are summarized in Tab. \ref{tab:requirements}. The whole detector is planned to be equipped with the same chip,
resulting in very stringent requirements on material budget, spatial resolution, detection efficiency, power density and fake hit rate. On the other hand, the expected radiation environment is relatively modest compared to the radiation levels expected in CMS or ATLAS at the same distance from the collision point. These requirements led to the choice of using Monolithic Active Pixel Sensors for the upgrade of the ITS.

\begin{table}[!t]
  \begin{center}
    \begin{tabular}{  l c  c  }
      \toprule
      \textbf{Parameter} & \textbf{Inner barrel} & \textbf{Outer barrel} \\ \midrule
      Silicon thickness & \multicolumn{2}{c }{ $50~\upmu$m}\\ 
      Spatial resolution & $5~\upmu$m & $10~\upmu$m\\ 
      Power density & $<300$ mW$/\text{cm}^2$ & $<100$ mW$/\text{cm}^2$\\ 
      Event resolution & \multicolumn{2}{c }{ $<30~\upmu$s}\\ 
      Detection efficiency & \multicolumn{2}{c }{ $>99\%$}\\ 
      Fake hit rate & \multicolumn{2}{c }{ $<10^{-5}$ hit per event per pixel}\\ 
      Average track density & $15$--$35$ $\text{cm}^{-2}$ & $0.1$--$1$ $\text{cm}^{-2}$\\ 
      TID radiation & $2700$ krad & $100$ krad\\ 
      NIEL radiation & $1.7 \times 10^{13}$ \NIEL & $10^{12}$ \NIEL \\ \bottomrule
    \end{tabular}
    \caption{Requirements for the Inner and Outer Barrel for the upgraded ITS. Radiation levels include a safety factor of $10$. Values are taken from \cite{ref:TDR}, but the radiation levels are updated according to the latest estimates.}
    \label{tab:requirements}
  \end{center}
\end{table}

%===================================%
\section{Technology choice}
\begin{figure}[!t]
  \begin{center}
    \includegraphics[width=0.6\textwidth]{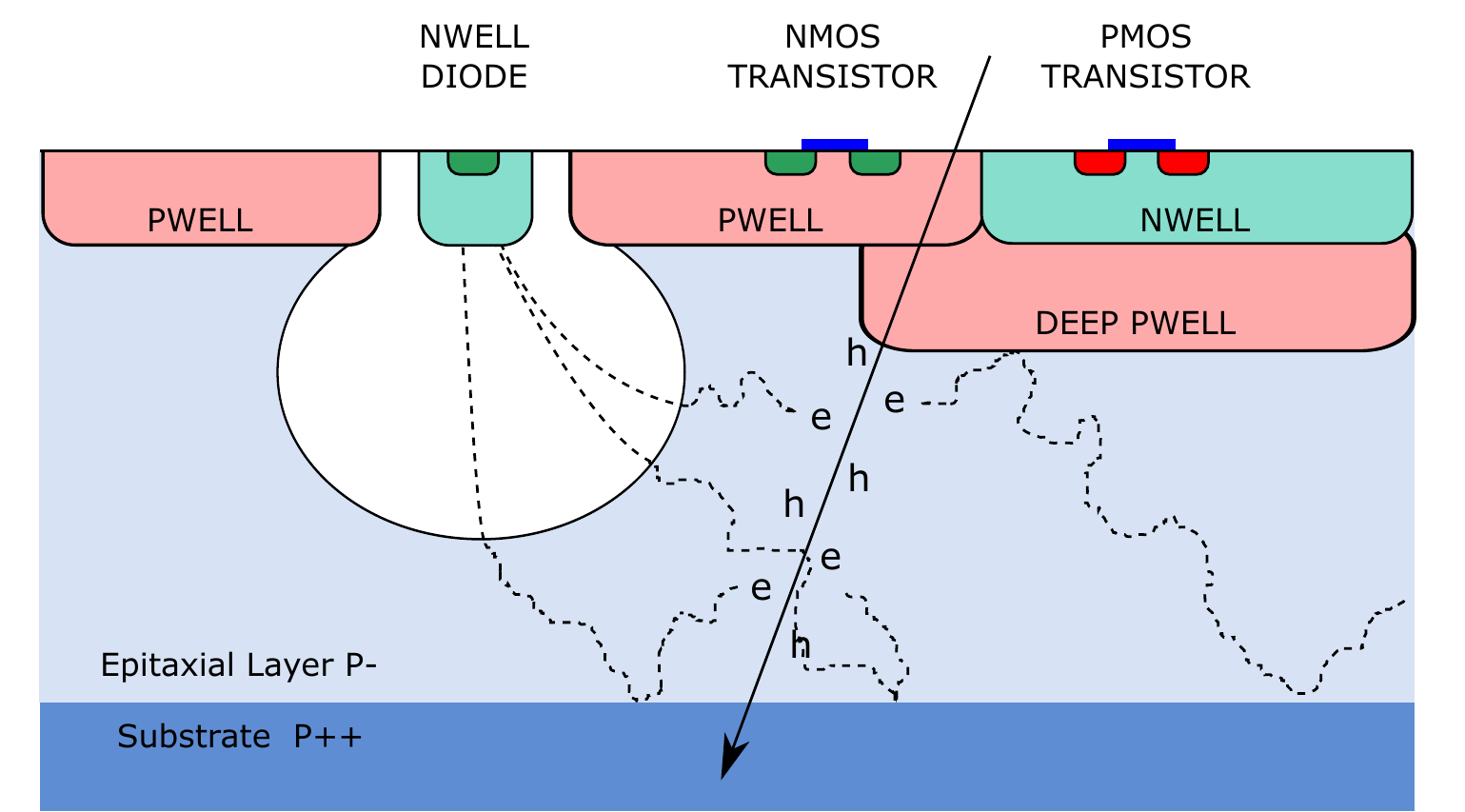}
    \caption{Cross section of one pixel in the TowerJazz $0.18~\upmu$m CMOS imaging process. Figure is taken from \cite{ref:TDR}.}
    \label{fig:technology}
  \end{center}
\end{figure}

  In the previous chapter the requirements for the chip for the ITS upgrade was discussed showing that the technology of CMOS Monolithic Active Pixel Sensors (MAPS) suits well these requirements. The first implementation of this technology in a heavy-ion physics experiment was the STAR PXL detector which has been installed in 2014 \cite{ref:STAR}. The ULTIMATE chip used for the STAR PXL detector however does not meet all the requirements of the ALICE ITS, particularly in terms of readout time. The ULTIMATE chip has a readout time of $190~\upmu$s whereas for the ITS upgrade the design goal is to stay below $30~\upmu$s. The development of a new chip is therefore needed for the ITS upgrade and the chosen technology is the TowerJazz $0.18~\upmu$m CMOS imaging process for which the schematic cross section of a pixel is shown in Fig. \ref{fig:technology}. The transistor feature size of $0.18~\upmu$m and the available six metal layers in this technology allow for the implementation of high density and low power digital circuitry. The small transistor feature size and the gate oxide thickness of $3~\upmu$m also make this technology more radiation tolerant than the $0.35~\upmu$m technology used up to now as the baseline in particle-physics applications. 

  The mechanism of the detection of particles in this technology is the following: the crossing particle generates electron-hole pairs in the high resistivity epitaxial layer; these electrons then diffuse until they reach the depleted volume within which they drift towards the collection n-well electrode where they are collected. Being able to produce the chip on a high resistivity epitaxial layer ($\gtrsim~1\text{ k}\Omega$~cm) is important, because it results in larger depletion volumes and thus better charge-collection efficiency. It also results in the electrons being less likely to be trapped, making the technology more radiation hard towards non-ionizing radiation. An important feature of the technology is the implementation of the deep p-well which allows the use of CMOS circuitry within the pixel area by shielding the n-well of the PMOS transistors from the epitaxial layer. Without such a shielding, the n-wells of the CMOS transistors would compete in the collection of the electrons with the collection diode and this would result in a lower charge-collection efficiency. For some prototypes a moderate bias voltage can be applied to the substrate ($< 10$ V) to create a larger depletion volume around the collection diode. This results in a smaller capacitance and consequently in a better charge-collection efficiency. The capacitance can be further reduced by keeping the collection n-well diode small, which in the case of the current prototypes is $2$--$3~\upmu$m.

\newpage
%===================================%
\section{Chip architectures}
\subsection{ALPIDE and MISTRAL-O}
\begin{figure}[!htbp]
  \begin{subfigure}[b]{0.49\textwidth}
    \begin{overpic}[width=1\textwidth]{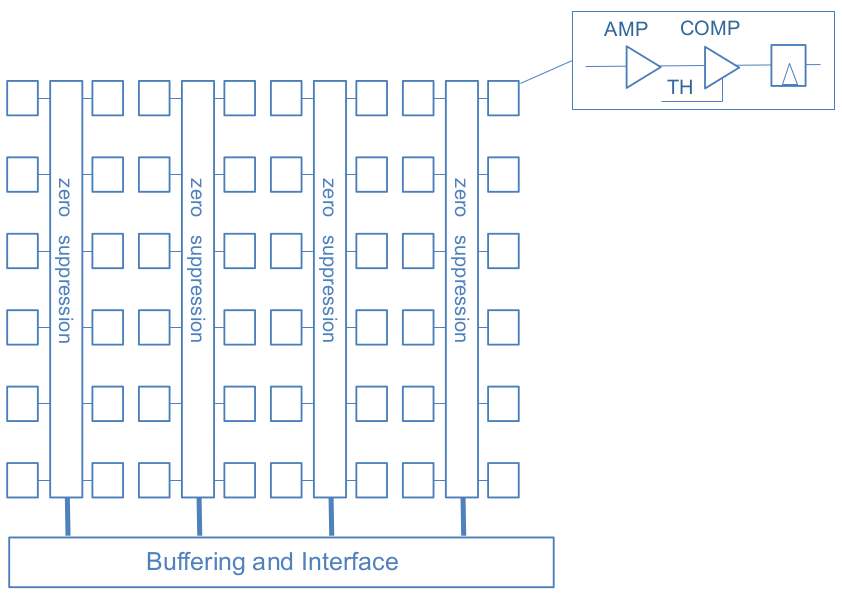}
      \put(   230,640){ALPIDE}
      \put( 650,300){\footnotesize Continuous or }
      \put( 650,250){\footnotesize external trigger}
    \end{overpic}
  \end{subfigure}\hspace{2.2em}
  \begin{subfigure}[b]{0.49\textwidth}
    \begin{overpic}[width=0.95\textwidth]{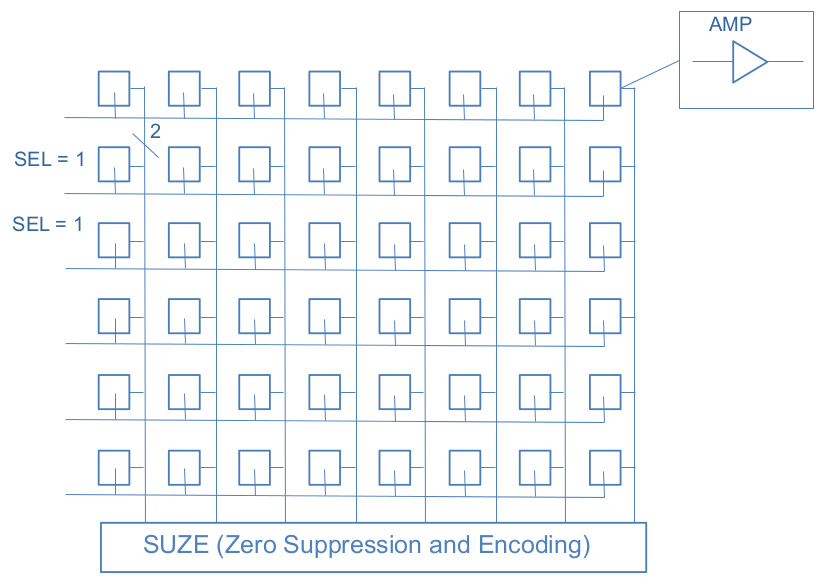}
      \put(   300,670){MISTRAL-O}
      \put( 800,300){\footnotesize Rolling }
      \put( 800,250){\footnotesize shutter}
    \end{overpic}
  \end{subfigure}
  \caption{Schematic design for the ALPIDE and MISTRAL-O architectures.}
  \label{fig:schematics}
\end{figure}
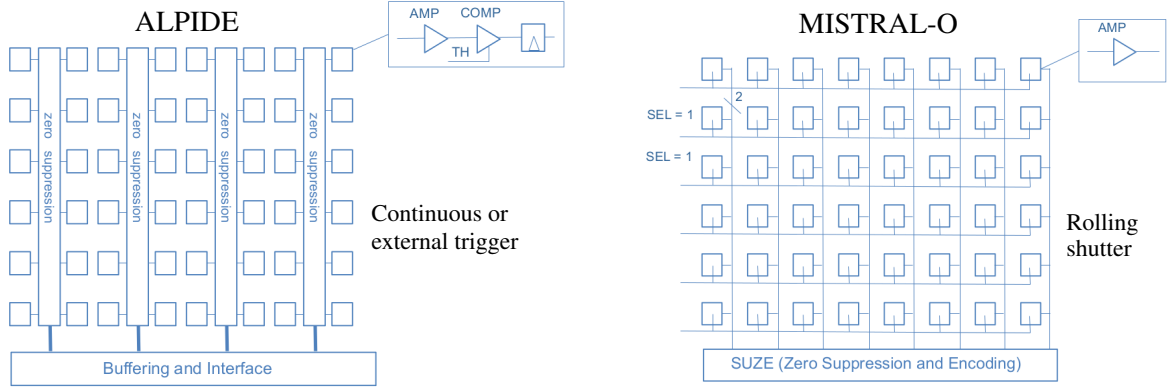

  There are two parallel chip developments ongoing for the upgrade of the ITS, the ALPIDE and the MISTRAL-O. The two chips are designed such that their dimensions, physical and electrical interface are identical, so either of them can be used for the upgrade without any modifications in the mechanical and electrical services. The schematic design of the matrix layout can be seen in Fig. \ref{fig:schematics} for the two architectures. 

  The MISTRAL-O development is a continuation of the ULTIMATE chip development. It has amplification within the pixel, the analog signal is then propagated to the end of column and the digitalization and zero suppression is done at the end of column. In the case of the ALPIDE the amplification, the digitalization and zero suppression is done within the pixel, so that only digital signals are propagated to the end of column. This allows for a data driven readout in the case of the ALPIDE while the MISTRAL-O is read out in the more traditional rolling shutter mode with an event-time resolution of around $20~\upmu$s. The event-time resolution of the ALPIDE is $\lesssim 2~\upmu$s and is determined by the rise time of the pulses. The power consumption, the dead area and the pixel pitch is smaller in the case of the ALPIDE (exact values can be seen in Tab. \ref{tab:ALPIDE_MISTRAL_comparison}) and both chips show good performance in noise occupancy and detection efficiency measurements which resulted in the choice of the ALPIDE as the baseline for the project. The current document focuses on the ALPIDE architecture, while details on the MISTRAL-O family can be found in \cite{ref:MistralTalk}.

\begin{table}[!htbp]
  \begin{center}
    \begin{tabular}{  l c  c  }
      \toprule
       & \textbf{ALPIDE} & \textbf{MISTRAL-O} \\ \midrule
      Pixel pitch & $28~\upmu$m $\times$ $28~\upmu$m & $36~\upmu$m $\times$ $65~\upmu$m \\ 
      Event time resolution & $\lesssim2~\upmu$s & $\sim{20}~\upmu$s \\ 
      Power consumption & $39\text{ mW}/\text{cm}^2$ & $80$--$90\text{ mW}/\text{cm}^2$ \\ 
      Dead area & $1.1$ mm $\times$ $30$ mm & $1.5$ mm $\times$ $30$ mm\\ \bottomrule
    \end{tabular}
    \caption{Performance and specification of the ALPIDE and MISTRAL-O architectures.}
    \label{tab:ALPIDE_MISTRAL_comparison}
  \end{center}
\end{table}

\subsection{pALPIDE-1}
\begin{figure}[!t]
  \begin{center}
    \begin{overpic}[width=0.45 \textwidth]{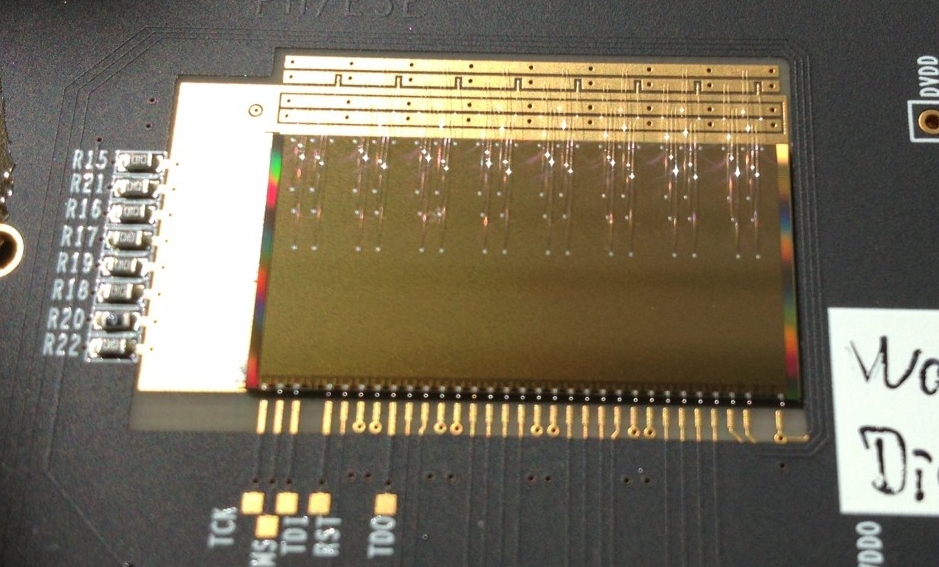}
      \put(475,93){\textcolor{white}{\large 30 mm}}
      \put(217,245){\rotatebox{85}{\large 15 mm}}
      \put(350,300){\textcolor{white}{0}}
      \put(470,298){\textcolor{white}{1}}
      \put(610,296){\textcolor{white}{2}}
      \put(750,294){\textcolor{white}{3}}
      \linethickness{1.5pt}
      \put(430,440){\color{white}\line(-0.05,-1){11.9}}
      \put(562,440){\color{white}\line(-0.03,-1){7.2}}
      \put(700,440){\color{white}\line(-0.013,-1){3.15}}
      \linethickness{0.9pt}
      \put(465,118){\color{white}\vector(-1,0.02){205}}
      \put(655,112){\color{white}\vector(1,-0.02){205}}
      \linethickness{0.7pt}
      \put(240,245){\vector(-0.1,-1){6}}
      \put(265,418){\vector(0.1,1){5.3}}
    \end{overpic}
    \caption{Picture of the pALPIDE-1 chip with the four sub-matrices indicated.}
    \label{fig:pALPIDE-1_photo}
  \end{center}
\end{figure}
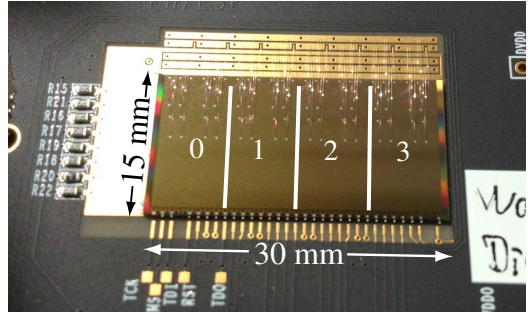

  The pALPIDE-1 chip is the first prototype of the ALPIDE family with the final chip size of $15$~mm~$\times$~$30$~mm. It has $512$ $\times$ $1024$ digital pixels with the size of $28~\upmu$m $\times$ $28~\upmu$m. It is read out using an asynchronous priority encoder network, thanks to which only hit pixels are read out resulting in very fast acquisition of the data from the full matrix. The chip is organized in four sub-matrices consisting of different types of pixels. A picture of the chip can be seen in Fig. \ref{fig:pALPIDE-1_photo} with these sub-matrices indicated. They differ in either the  reset mechanism of the pixels or in the spacing between the collection diode and the surrounding p-well. In the following, results will be reported for only one of these pixel variations. 

%===================================%
\section{Characterization of the pALPIDE-1 prototype}
\subsection{Methods and tools}
\begin{figure}[!htbp]
  \begin{subfigure}[b]{0.49\textwidth}
  \begin{center}
    \begin{overpic}[width=0.9\textwidth]{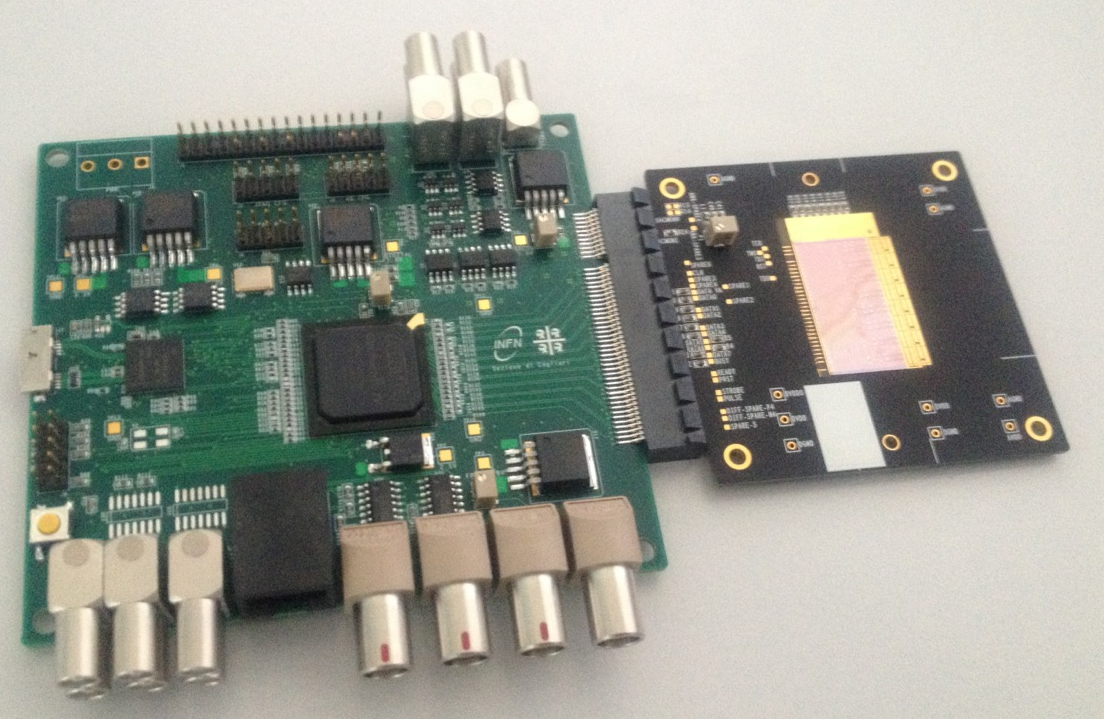}
      \put(70,560){DAQ board}
      \put(700,160){Carrier card}
      \put(600,550){pALPIDE-1}
      \linethickness{1.5pt}
      \put( 690,540){\vector(0.5,-1){80}}
    \end{overpic}
    \caption{Photo of the readout board and the chip wire bonded to the carrier card.}
    \label{fig:carrier}
  \end{center}
  \end{subfigure}
  \begin{subfigure}[b]{0.49\textwidth}
    \begin{center}
      \begin{overpic}[width=0.85\textwidth]{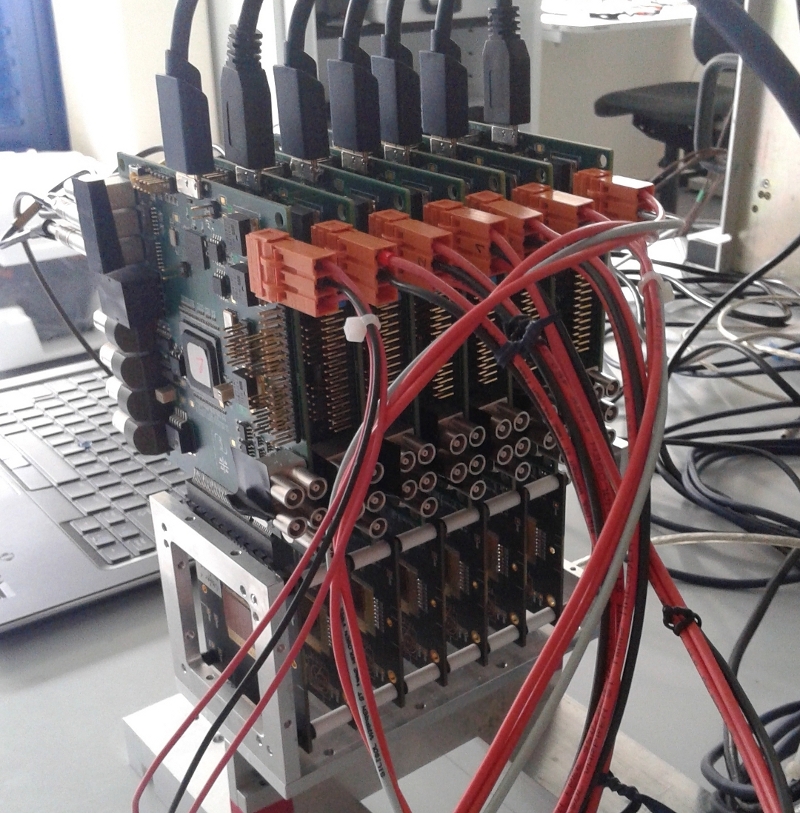}
        \linethickness{2pt}
        \put(100,125){\color{blue}\vector(2,0.77){700}}
        \put(70,135){\color{blue}\rotatebox{20}{Beam}}
        \put(380,160){\color{white}\rotatebox{22}{7 pALPIDE-1}}
        \put(380,500){\color{white}\rotatebox{20}{7 DAQ boards}}
      \end{overpic}
      \caption{Test beam setup.}
      \label{fig:testBeamSetup}
    \end{center}
  \end{subfigure}
  \caption{Setup used in the measurements for the readout of the pALPIDE-1.}
  \label{fig:testBeam}
\end{figure}
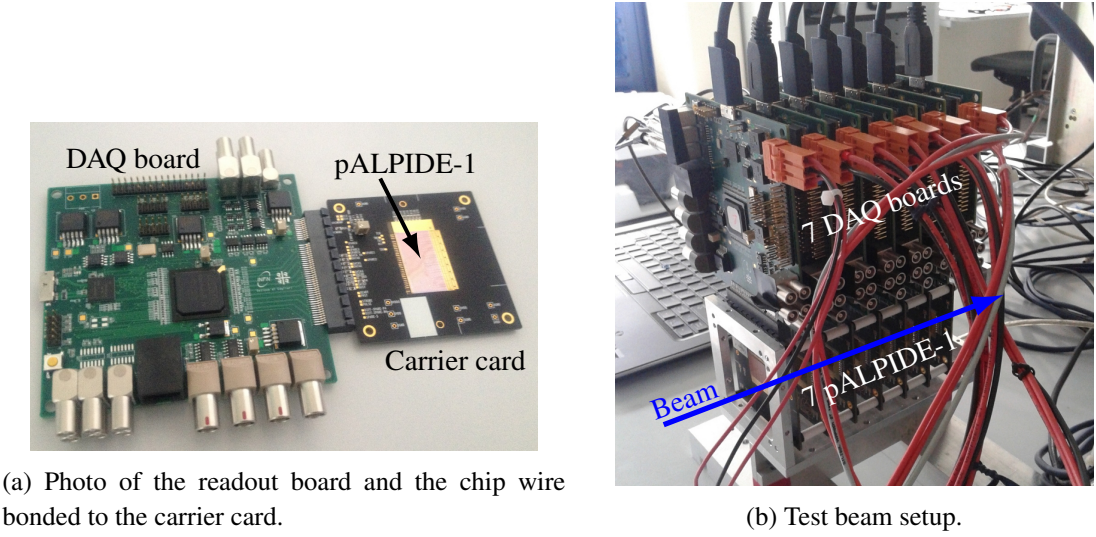

  In the upgraded ITS the pixel chips will be laser soldered to the flexible printed circuit which will give both the mechanical and the electrical connection, while for testing, the prototypes are wire bonded to a carrier card as can be seen in Fig. \ref{fig:pALPIDE-1_photo}. The carrier card is then connected to the readout board by a PCI Express connector and the readout board is connected via USB to the computer (Fig. \ref{fig:carrier}). The first tests of the chips (noise and threshold measurements, noise occupancy determination) are done in this setup while for the detection efficiency and spatial resolution measurements a stack of this setup is used. The stack consists of seven pALPIDE-1 chips where the outer six are the tracking planes and the central one is being tested (Fig. \ref{fig:testBeamSetup}). The seven chips are connected to a computer the same way as the individual chips are and the readout is done by the EUDAQ framework \cite{ref:eudaq}. The analysis of the data collected at beam tests is then done by the EUTelescope software \cite{ref:eutelescope}.

\subsection{Performance}
  The pALPIDE-1 prototype has been thoroughly characterized both in laboratory measurements and in several test beam campaigns at different test beam facilities (PS, SPS, DESY, BTF and PAL). The beam energy at these facilities range from $60$ MeV to $120$ GeV and the particles used are electrons, positrons and pions. The following results were obtained at the PS using a $6$~GeV negative pion beam.

\begin{figure}[!htbp]
  \begin{subfigure}[b]{0.48\textwidth}
    \begin{center}
      \includegraphics[width=1\textwidth]{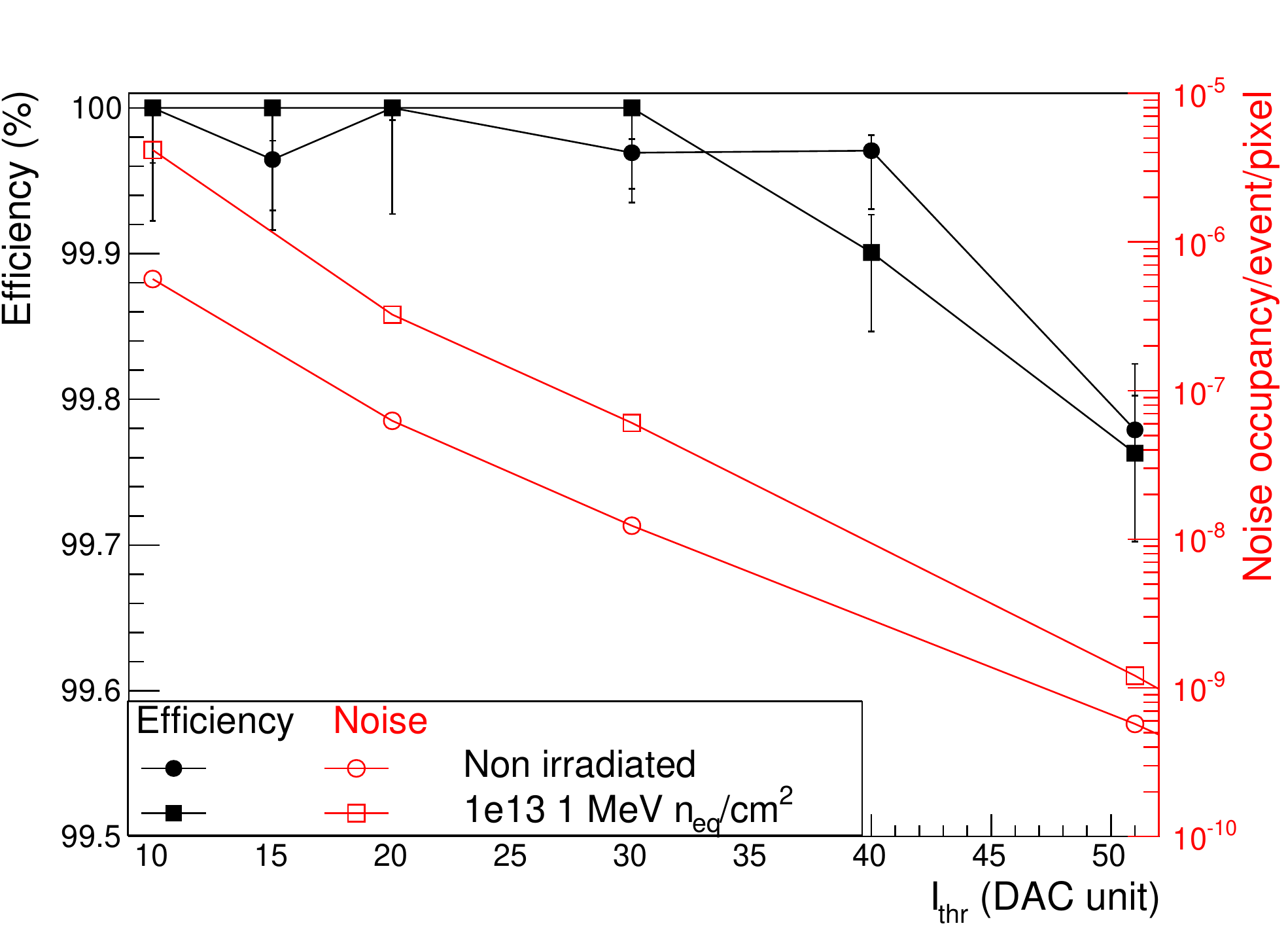}
      \caption{Detection efficiency (left axis) and noise occupancy (right axis) results. The noise occupancy values were calculated after masking the $20$ most noisy pixels.}
      \label{fig:efficiency_noiseOccupancy}
    \end{center}
  \end{subfigure}\hspace{1.4em}
  \begin{subfigure}[b]{0.48\textwidth}
    \begin{center}
      \includegraphics[width=1\textwidth]{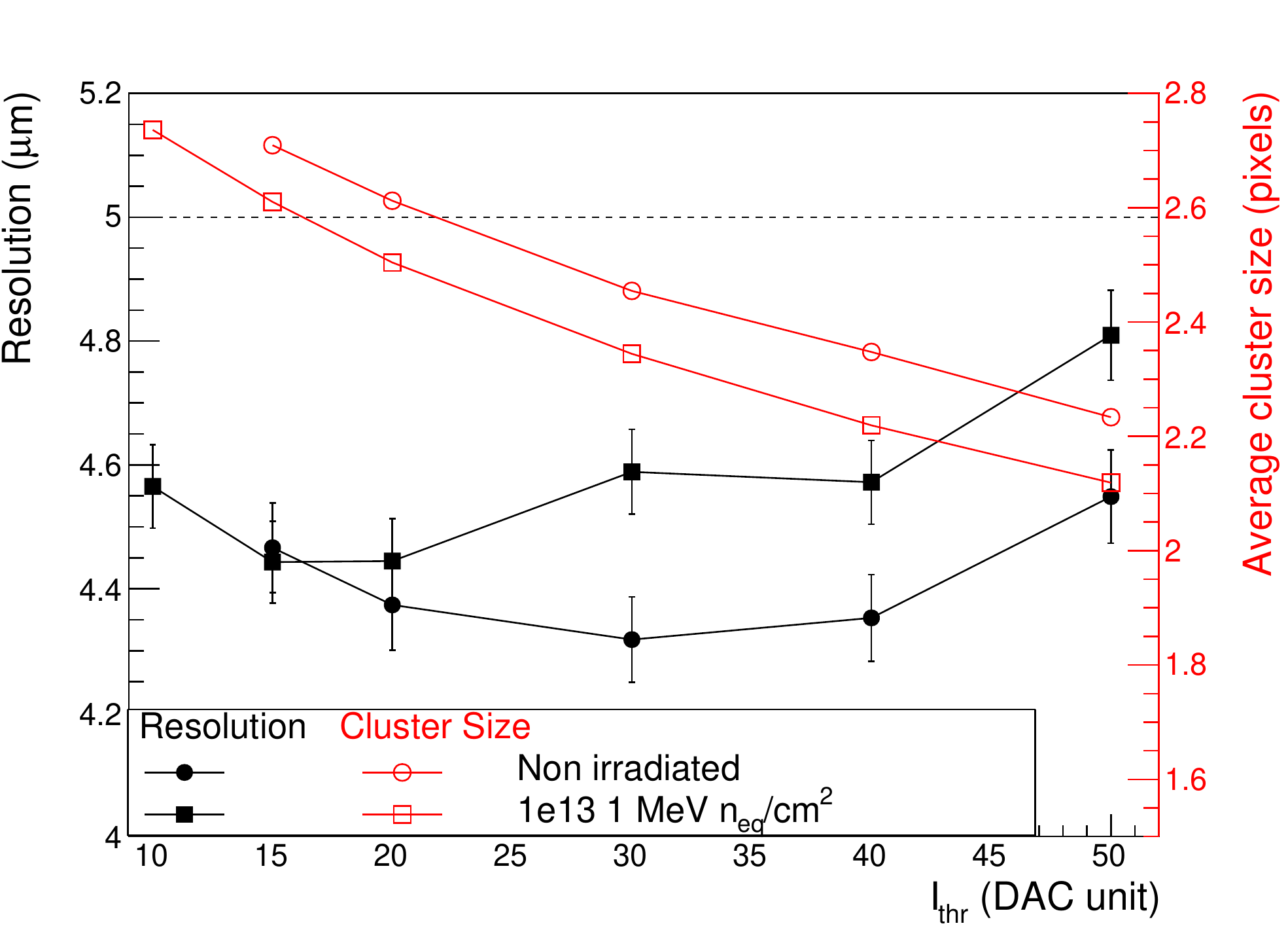}
      \caption{Spatial resolution (left axis) and average cluster size (right axis) results. The spatial resolution is calculated by assuming $2.39~\upmu$m pointing resolution for the tracks.}
      \label{fig:resolution_clusterSize}
    \end{center}
  \end{subfigure}
  \caption{Results for the pALPIDE-1 from measurements at the PS using a $6$ GeV negative pion beam. The data for both figures were taken applying $-3$ V reverse substrate bias and is shown as function of the threshold current.}
  \label{fig:results}
\end{figure}

  The results are summarized in Fig. \ref{fig:results} where the left panel shows the detection efficiency and noise occupancy values of the pALPIDE-1 while the right panel shows the spatial resolution and the average cluster size of the chip. Both figures show the results as a function of the threshold current which is a parameter of the chip used for setting the input charge threshold of the pixels. The requirements for the ITS upgrade is to have a detection efficiency better than $99\%$ and noise occupancy lower than $10^{-5}$ hits/event/pixel. In Fig. \ref{fig:efficiency_noiseOccupancy} it can be seen that both requirements can be fulfilled at a wide operating range of the threshold current. The performance in terms of detection efficiency does not degrade after irradiating the chip by $10^{13}$ \NIEL, but the noise occupancy values become slightly higher. In Fig. \ref{fig:resolution_clusterSize} it can be seen that there is also a large operational margin in terms of spatial resolution where the requirement for the chip is to stay below $5~\upmu$m which is fulfilled both before and after irradiation by $10^{13}$ \NIEL.

\begin{figure}[!t]
  \begin{subfigure}[b]{0.46\textwidth}
    \begin{center}
      \begin{overpic}[width=1.05\textwidth]{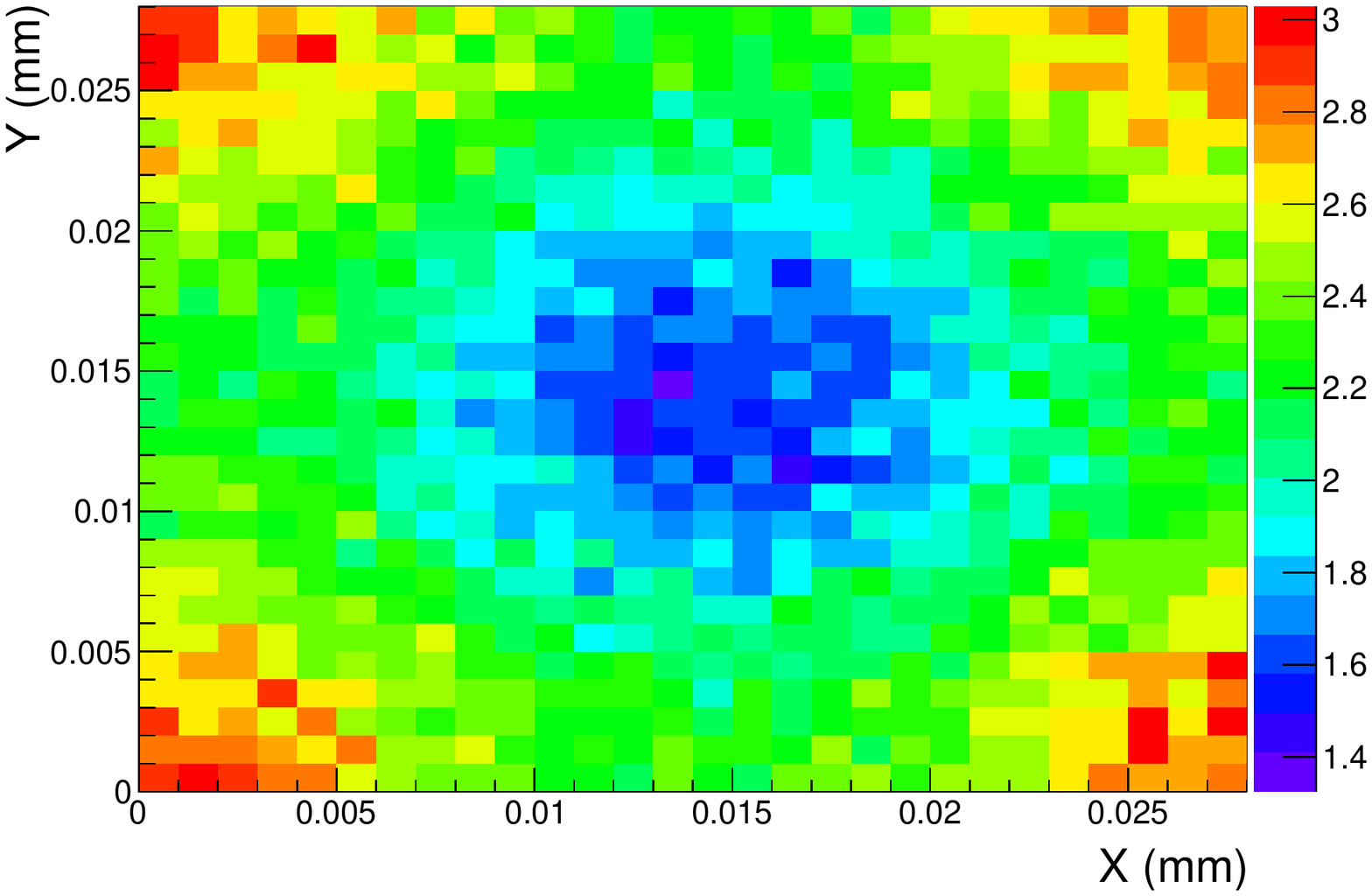}
        \linethickness{2pt}
        \put(320,710){Cluster size vs track}
        \put(240,650){impinging point within pixel}
        \put(100,70){\vector(0,1){270}}
        \put(100,340){\vector(1,0){440}}
        \put(540,340){\vector(-1,-0.62){450}}
        \put(40,20){\huge 1}
        \put(40,340){\huge 2}
        \put(500,340){\color{white}{\huge 3}}
      \end{overpic}
      \caption{Dependency of the average cluster size on the impinging point of the track within the pixel.}
      \label{fig:clusterSize2D}
    \end{center}
  \end{subfigure}\hspace{3em}
  \begin{subfigure}[b]{0.46\textwidth}
    \begin{center}
      \begin{overpic}[width=0.95\textwidth]{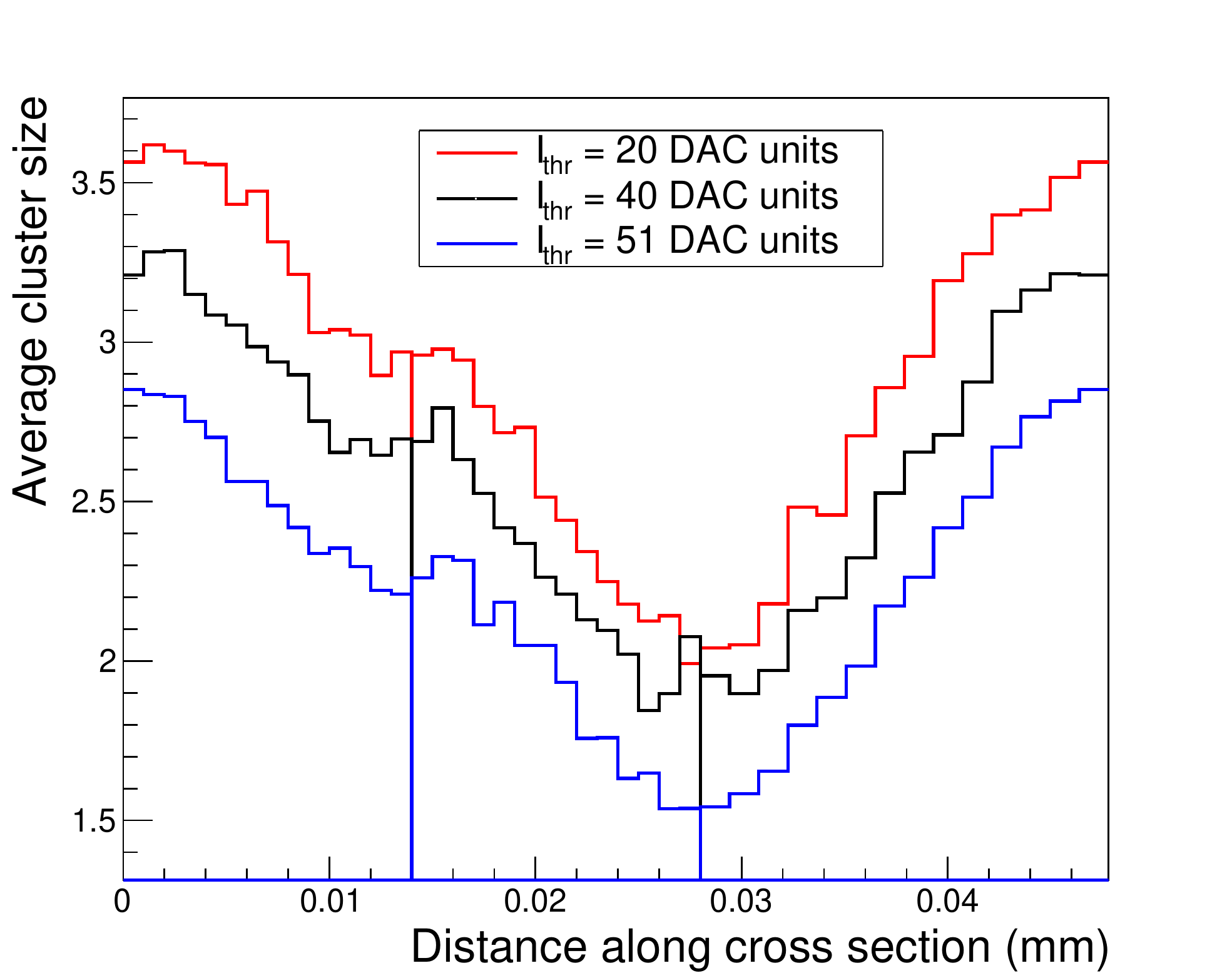}
        \put(320,730){Cross section}
        \put(100,150){\huge 1}
        \put(300,150){\huge 2}
        \put(550,150){\huge 3}
        \put(850,150){\huge 1}
      \end{overpic}
      \caption{Cross section of figure (\subref{fig:clusterSize2D}) along the indicated arrows.}
      \label{fig:clusterSize2D_CS}
    \end{center}
  \end{subfigure}
  \caption{Results for the pALPIDE-1 from measurements at the PS using a $6$ GeV negative pion beam. The data for both figures were taken with non optimal setting for the chip (applying $0$ V reverse substrate bias) to emphasize the effect on the cluster size.}
  \label{fig:clusterSize2D_full}
\end{figure}

  In Fig. \ref{fig:clusterSize2D_full} the average cluster size is shown as a function of the impinging point of the track within a pixel. It is smallest if the track goes through the center of the pixel, close to the collection diode and largest if the track goes through the corner of the pixels where it is more likely to have charge-sharing with the neighboring pixels.

%===================================%
\section{Summary}
  The Inner Tracking System of ALICE will be upgraded during the second long shut down of the LHC in 2019--2020 to fulfill the physics requirements of ALICE after the shut down. The new detector will have seven layers of Monolithic Active Pixel Sensors and will improve significantly the tracking and \pT{} resolution at low \pT{}, as well as the impact-parameter resolution. The first large-scale prototypes of both ALPIDE and MISTRAL-O have been thoroughly characterized and show very good performance with a wide operational margin. Tests have also been carried out with irradiated sensors and the chips show good performance after an irradiation level of $10^{13}$~\NIEL. Overall the development of the new detector is progressing according to schedule with all aspects of the R\&D close to completion.

%===================================%
\section{Acknowledgement}
We would like to thank the host institutes (DESY, BTF Frascati, Pohang and CERN) for their support during our test beam measurements. We would like to thank also the developers of the EUDAQ and EUTelescope software framework for their support.

\end{document}